# Interlayer Engineering of Lattice Dynamics and Elastic Constants of 2D Layered Nanomaterials under Pressure

Guoshuai Du,[1,2,#] Lili Zhao,[3,#] Shuchang Li,[4,#] Jing Huang,[5] Susu Fang,[3] Wuxiao Han,[1,2] Jiayin Li,[1,6] Yubing Du,[1,2] Jiaxin Ming,[1,6] Tiansong Zhang,[1,2] Jun Zhang,[7] Jun Kang,[5,*] Xiaoyan Li,[4,*] Weigao Xu,[3,*] and Yabin Chen[1,2,8,*]

[1]*Advanced Research Institute of Multidisciplinary Sciences (ARIMS), Beijing Institute of Technology, Beijing 100081, China*

[2]*School of Aerospace Engineering, Beijing Institute of Technology, Beijing 100081, China*

[3]*School of Chemistry and Chemical Engineering, Nanjing University, Nanjing 210023, China*

[4]*Mechano-X Institute, Applied Mechanics Laboratory, Department of Engineering Mechanics, Tsinghua University, Beijing 100084, China*

[5]*Beijing Computational Science Research Center, Beijing 100193, China*

[6]*School of Chemistry and Chemical Engineering, Beijing Institute of Technology, Beijing 100081, China*

[7]*Institute of Semiconductors, Chinese Academy of Sciences, Beijing 100083, China*

[8]*BIT Chongqing Institute of Microelectronics and Microsystems, Chongqing 400030, China*

[#]These authors contributed equally to this work.
[*]Correspondence and requests for materials should be addressed to J.K. (jkang@csrc.ac.cn), X.L. (xiaoyanlithu@tsinghua.edu.cn), W.X. (xuwg@nju.edu.cn), and Y.C. (chyb0422@bit.edu.cn).

# ABSTRACT

Interlayer coupling in two-dimensional (2D) layered nanomaterials can provide us novel strategies to evoke their superior properties, such as the exotic flat bands and unconventional superconductivity of twisted layers, the formation of moiré excitons and related nontrivial topology. However, to accurately quantify interlayer potential and further measure elastic properties of 2D materials remains vague, despite significant efforts. Herein, the layer-dependent lattice dynamics and elastic constants of 2D nanomaterials have been systematically investigated via pressure-engineering strategy based on ultralow frequency Raman spectroscopy. The shearing mode and layer-breathing Raman shifts of $MoS_2$ with various thicknesses were analyzed by the linear chain model. Intriguingly, it was found that the layer-dependent $d\omega/dP$ of shearing and breathing Raman modes display the opposite trends, quantitatively consistent with our molecular dynamics simulations and density functional theory calculations. These results can be generalized to other van der Waals systems, and may shed light on the potential applications of 2D materials in nanomechanics and nanoelectronics.

# INTRODUCTION

Van der Waals interactions between the atomic layers facilitate us to mechanically exfoliate and transfer various two-dimensional (2D) layered nanomaterials, and further to controllably stack their heterostructures.[1-5] Despite its relatively weak characteristic, van der Waals interactions play an essential role in the exotic physical properties of these quantum systems, such as twist angle-dependent moiré superlattices, the prominent spin-orbital couplings in 2D heavy fermions, and the correlated topological insulators.[6-9] Therefore, the quantitative determination and modulation of interlayer coupling strength in 2D materials are of significant importance for us to boost their practical applications in twistronics and optoelectronics, however, it is with huge challenge.

Previously, the interlayer van der Waals interactions in transition metal dichalcogenides have been tentatively quantified according to pressure-enhanced valance band splitting.[10] Obviously, this approach had material-specific prerequisite, severely relying on 2D samples displaying the noticeable splitting energy in their valance bands. In addition, the interlayer strength of van der Waals pressure has been measured by trapping pressure-sensitive molecules between graphene layers.[11] Nevertheless, these organic molecules themselves can spontaneously induce the efficient charge transfer, and consequently influence the intrinsic interlayer interactions of 2D materials. Moreover, a synchrotron X-ray diffraction experiment has been established to quantify the interlayer interactions in $TiS_2$, and an unexpected accumulation of electron density in the interlayer region was found,[8] which is difficult to popularize due to the complex apparatus. The surface indentation method could measure the perpendicular-to-the-plane elasticity or the adhesion forces between probe tip and 2D nanoflake, while the results should be evaluated against benchmarks rather than treated as standalone measurements of adhesion energy.[12,13] In response to these limitations, a strategic approach based on ultralow frequency (ULF) Raman spectroscopy has been developed to quantify the interlayer interactions in multilayer graphene.[14] Such non-invasive, rapid and convenient technique can be widely extended to numerous 2D materials, and it is suitable for diverse experimental environments as well.

Hydrostatic pressure can effectively modulate the lattice structures and physical properties of 2D nanomaterials, without deliberately introducing any impurities or defects.[15] Extreme pressure can induce interlayer slippage, structural evolutions and metal-insulator transitions in 2D materials and heterostructures, owing to their pressure-sensitive interlayer and intralayer interactions.[15-18]

Typically, interlayer interactions response more sensitively on pressure than intralayer interactions governed by strong covalent bond. At the same time, the interlayer coupling strength of 2D layered nanomaterial strictly depends on its layer number ($N$).[19] For instance, diamond anvil cell (DAC) combined with synchrotron X-ray diffraction has been utilized to investigate the evolutional mechanism and stability of bulk $MoS_2$, and the obtained equation of state suggested its bulk modulus of bulk $MoS_2$ as ~79.5 GPa, while the modulus of ultrathin $MoS_2$ remains uncertain.[20] As the thickness reduced to monolayer limit, high pressure can principally tune its electronic energies and band structure, such as the transition from direct to indirect band gap.[21,22] Additionally, the performance of $MoS_2$ nanotransistors can be notably improved under hydrostatic pressure.[23] Lattice vibration and structural stability of ultrathin $MoS_2$ were initially studied under high pressure, implying that both interlayer and intralayer force constants dramatically changed with pressure.[22,24-29] To the best of our knowledge, the comprehensive studies on the pressure-regulated lattice dynamics and elastic constants of $MoS_2$ with various $N$ is still lacking, and the associated mechanism remains unclear.

In this work, pressure engineering approach were developed to systematically investigate the layer-number dependent interlayer interactions in 2D layered materials using *in situ* ULF Raman spectroscopy, by taking $MoS_2$ as the typical paradigm. Our experimental results were analyzed with the linear chain model and consistent with the density functional theory (DFT) calculations and molecular dynamics (MD) simulations as well. It is found that force constants and modulus of $MoS_2$ were dramatically enhanced with pressure, and the evolution of interlayer interactions exhibited a clear layer number (thickness) dependence. The results in this study can provide a pivotal reference to clarify the servo response of mechanical properties of 2D materials under extreme pressure.

**RESULTS AND DISCUSSION**

Here we took the commonly studied 2$H$-$MoS_2$ (space group $P6_3/mmc$) with layered structure as a paradigmatic example. In principle, the relevant results can be rationally generalized to a wide family of van der Waals materials and their heterostructures. According to the lattice dynamics theory, each primitive unit cell of $MoS_2$ consists of two molecules,[24] that is, two Mo atoms and four S atoms. Hence, there exists 18 normal vibrational modes at the Brillouin zone center with the irreducible representations as $\Gamma = 2A_{2u} \oplus 2E_{1u} \oplus E_{2u} \oplus B_{1u} \oplus A_{1g} \oplus 2B_{2g} \oplus E_{1g} \oplus 2E_{2g}$, where except for one $A_{2u}$ and one $E_{1u}$ acoustic modes, the remained ones originate from the long-wavelength optical phonons.[30] In experiment, three Raman peaks of bulk MoS$_2$, *i.e.*, low frequency $E_{2g}^2$ at 32 cm$^{-1}$ and high-frequency $E_{2g}^1$ (383 cm$^{-1}$) and $A_{1g}$ (408 cm$^{-1}$), are usually observed without considering the resonance effect.[31] Importantly, the $E_{2g}^2$ (shear mode (S)), and in-plane $E_{2g}^1$ and out-of-plane $A_{1g}$ modes vibrate along the interlayer and intralayer directions, respectively, as shown in Figure 1a. Notably, the intralayer vibrations refers to the relative motions of atoms within an individual layer, whereas the interlayer vibrations are the collective excitations of the entire MoS$_2$ layer perpendicular or parallel to the normal direction. More interestingly, these vibrational frequencies are principally sensitive to the *N*.[32] When the thickness of bulk 2*H*-MoS$_2$ is shrank to few layer, the point group $D_{6h}^4$ is changed to $D_{3d}$ (even number) or $D_{3h}$ (odd number), as shown in Note 1.[33] For even *N*, its layer breathing (LB) and S modes are classified as either Raman-active or IR-active; for odd *N*, the LB modes can be either Raman-active or IR-active, while the S modes can be Raman-active or both Raman-active and IR-active. In normal, *N*-layer MoS$_2$ typically includes (*N*-1) two-fold degenerated *S* modes and (*N*-1) nondegenerate LB modes, as shown in Figures S1 and S2.

Figure 1b displays the representative few-layer MoS$_2$ nanoflake exfoliated from the high-quality bulk crystal, with the thickness exactly determined by the combined optical contrasts and low-frequency Raman spectrum. Figure 1c presents the detailed stokes and anti-stokes Raman spectra of MoS$_2$ nanoflake with various *N*, acquired using our homemade Raman system with its layout shown in Figure S3. The extracted *N*-dependent Raman shifts are summarized in Figure 1d. As *N* increases, the high-frequency $E_{1g}^1$ and $A_{1g}$ modes shift oppositely till saturation when *N* exceeds around 5, well consistent with the literature.[31]

Next, for the low-frequency $E_{2g}^2$ and $B_{2g}$ modes, the monoatomic chain model (MCM) can be used to explain the experimental results, when considering each individual layer rigidly vibrates against its adjacent. In this scenario, each MoS$_2$ layer can be simplified as a rigid ball with the mass density $\mu$ (the total mass per unit area), and interlayer interactions can be represented by the elastic constant $\alpha$ between balls. Therefore, by diagonalizing the $N \times N$ dynamical matrix, the frequencies of (*N*-1) two-fold degenerate S modes and (*N*-1) LB modes can be calculated as $\omega(S_{N,N-j})=\omega(S_{bulk})\sin(j\pi/2N)$, and $\omega(LB_{N,N-j})=\omega(LB_{bulk})\sin(j\pi/2N)$, respectively,[33] where the $\omega(S_{bulk})=\sqrt{\alpha^{\parallel}/\mu}/\pi c$ and $\omega(LB_{bulk})=\sqrt{\alpha^{\perp}/\mu}/\pi c$ are attributed from the S and LB modes of bulk crystals and $j=1, 2, 3, …, N-1$. $\alpha^{\parallel}$ ($\alpha^{\perp}$) corresponds to the in-plane (out-of-plane) interlayer force constant per unit area between the nearest-neighbors, and $c$ is the speed of light. Obviously, $S_{N,1}/LB_{N,1}$ with $j = N-1$ is the highest-frequency branch, while $S_{N, N-1}/LB_{N, N-1}$ ($j=1$) means the lowest-frequency branch. As shown in Figure 1d, our experimental data can be well fitted with the MCM, implying the negligible coupling of MoS$_2$ layer with its substrate. Besides, the third highest-frequency $S_{N,3}$ mode together with the lowest-frequency $LB_{N, N-1}$ and the third lowest-frequency $LB_{N, N-3}$ modes of few layer MoS$_2$ have been observed in Raman spectrum as well, despite their weak intensities.[34] The $\alpha^{\parallel}$ and $\alpha^{\perp}$ for 2$H$-MoS$_2$ under ambient condition were derived as $2.8\times10^{19}$ and $8.6\times10^{19}$ N m$^{-3}$, respectively, which is quantitatively consistent with those in literature.[34] Furthermore, we performed the detailed DFT calculations and MD simulations to gain more insights on the $N$-dependence of vibrational frequency of MoS$_2$, and the results were in good agreement with the experimental data, as illustrated in Figure S4.

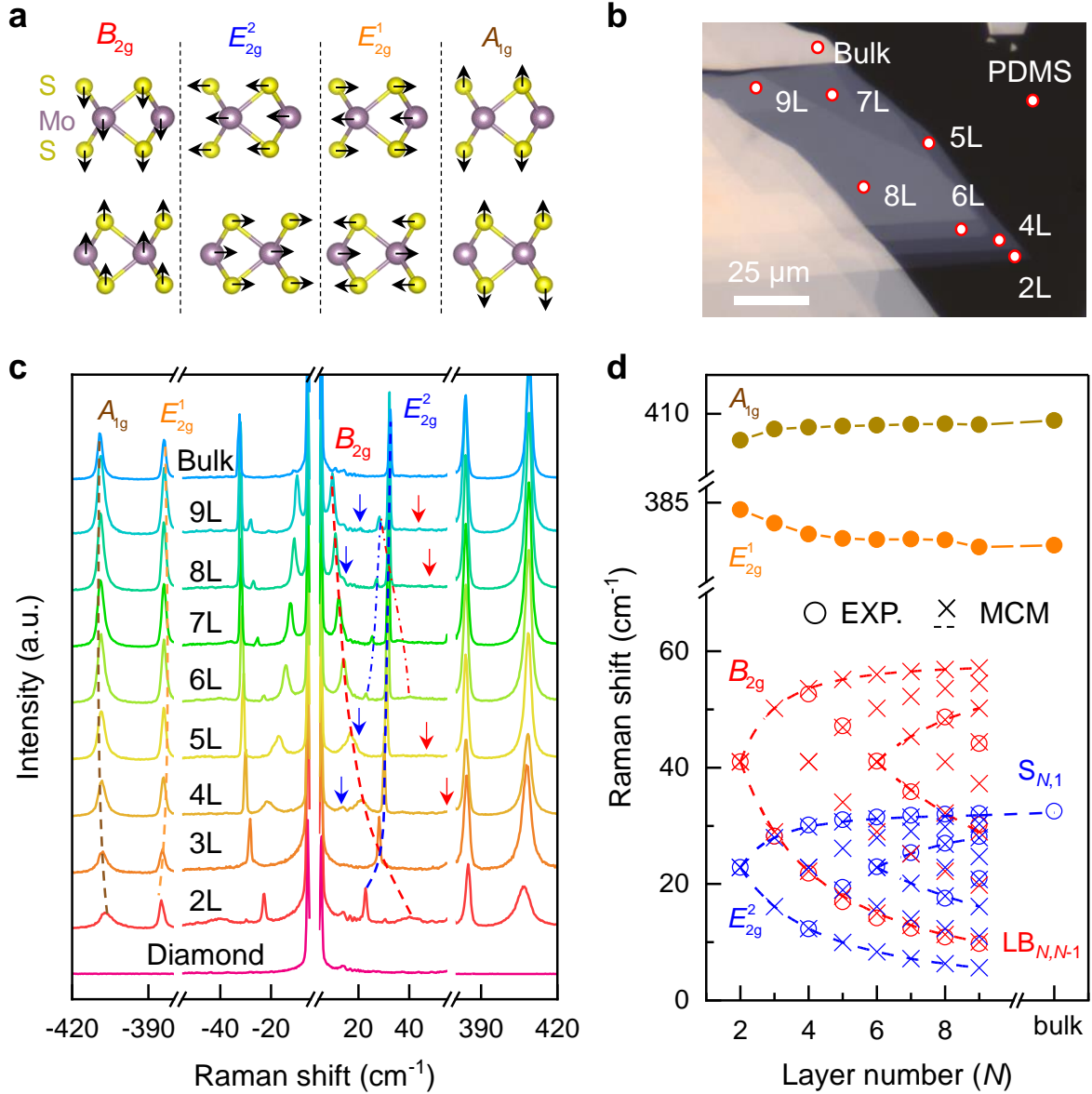

**Figure 1. Collective LB and S modes as well as high-frequency intralayer Raman modes of MoS$_2$ under ambient pressure. a)** Scheme of Raman-active normal modes of 2D layered MoS$_2$, including the collective LB ($B_{2g}$) and S ($E_{2g}^2$) modes, and intralayer $E_{2g}^1$ and $A_{1g}$ modes. **b)** Representative optical image of the exfoliated MoS$_2$ on a polydimethylsiloxane (PDMS) surface, and the layer numbers of MoS$_2$ are labeled. **c)** Stokes and anti-Stokes Raman spectra of the few-layer and bulk MoS$_2$ on a diamond surface. The dashed lines and arrows are used for guide. **d)** Raman shifts of few-layer MoS$_2$ as a function of $N$. The open circles denote the experimental data. The blue (S mode) and red (LB mode) crosses mean the obtained calculation results based on the MCM, and the data on each dashed line are from the same branch.

To probe the pressure effect on the interlayer and intralyer couplings of MoS$_2$, we transferred the few-layer MoS$_2$ onto the diamond culet and the extreme hydrostatic pressure can be generated with a symmetrical DAC, as shown in Figure 2a. Such a GPa-level pressure can conveniently shorten both the interlayer and intralayer distances, and thus effectively modulate the relevant Raman frequencies. Figure 2b exhibits the evolution of Raman spectrum of 6L MoS$_2$ under high pressure up to 9.6 GPa, and the structural transitions may happen at such high pressure.[22] From our MD simulations, four low-frequency Raman modes as S$_{6,1}$, S$_{6,3}$, LB$_{6,3}$ and LB$_{6,5}$ can be detected under ambient pressure (Figure S5), as discussed previously in Figure 1c. When subjected to high pressure, the LB$_{6,3}$ mode weakened dramatically and became undetectable in our experiments (Figure 2b). The intensity of three remained modes gradually decreased as the pressure increased, primarily owing to the phonon scattering with the surrounding transmitting medium or the enriched charge carriers. Under high pressure, all Raman frequencies of S$_{6,1}$, S$_{6,3}$ and LB$_{6,5}$ exhibit the significant blueshift, which can be well explained by the compressed lattice constants and resultant enhanced lattice structure of MoS$_2$ under pressure. Typically, the lattice constant *c* is reduced by 6.4% at ~10 GPa.[20] Impressively, the S modes strengthened more raplidly with a larger slope d$\omega$/d*P* than LB mode in Figure 2c. In general, LB modes involve the out-of-plane atomic motions and should be therefore more sensitive to the variation of interlayer spacing than S modes. However, the intuitive analysis is indeed inconsistent with our experimental data in Figure 2c. Below we will make the detailed discussions on this unexpected phenomenon. On the other hand, the full width at half-maximum (FWHM) of each peak became larger under high pressure, and the LB mode presents more prominent, indicating its shorter lifetime $\tau$, due to the anharmonic decay of phonon transport. For instance, the FWHM of LB$_{6,5}$ mode was enlargred from ~2.3 cm$^{-1}$ (0 GPa) to 9.5 cm$^{-1}$ (8 GPa), resulting in the reduced $\tau$ from 14.3 to 3.5 ps based on the uncertainty principle $\Delta E \sim \hbar/\tau$.

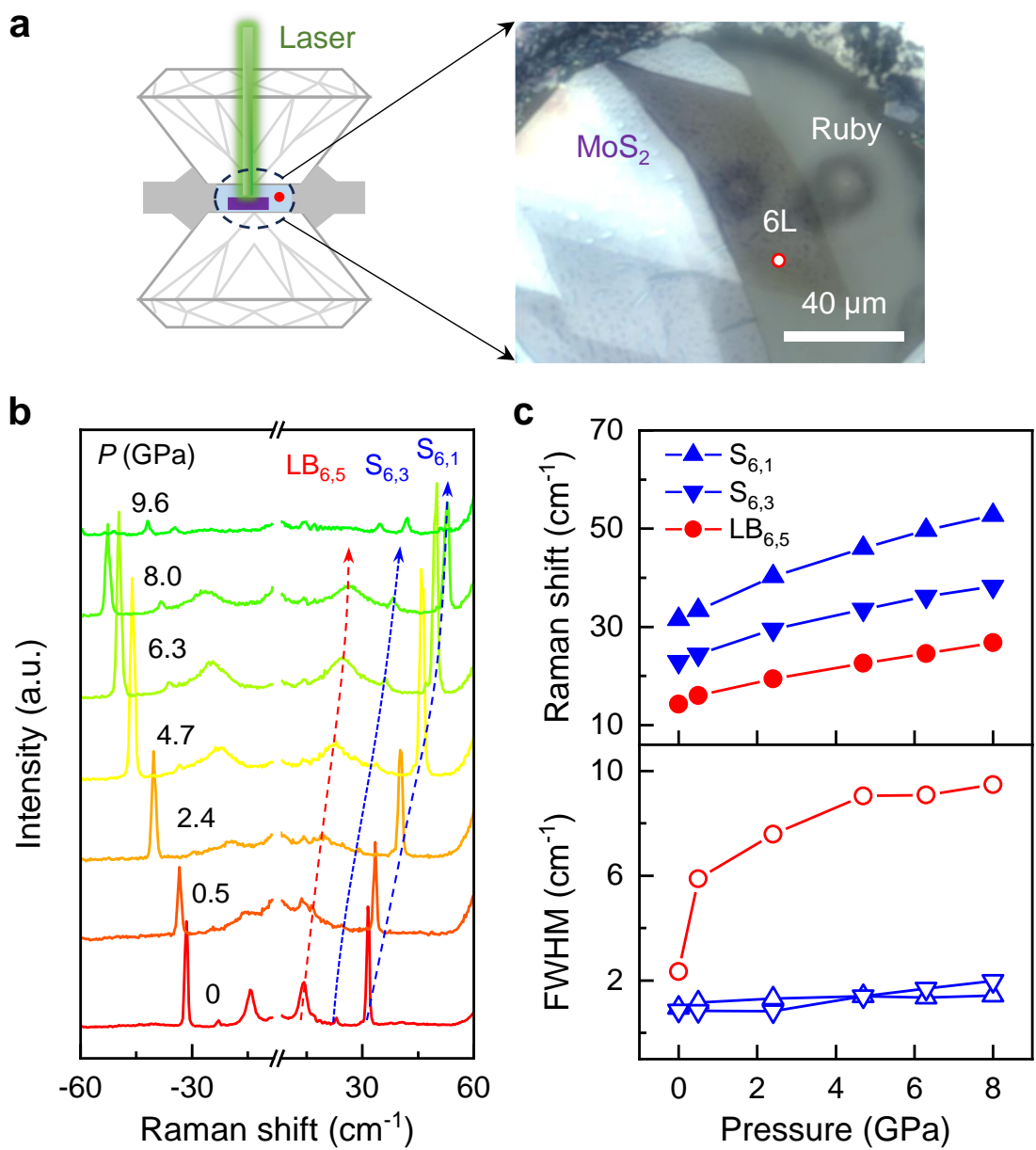

**Figure 2. Low-frequency LB and S modes of 6L-MoS$_2$ under high pressure applied by a DAC. a)** MoS$_2$ nanoflakes transferred onto diamond surface. The mixture of methanol-ethanol-water (16:3:1) was used as the pressure transmission medium, and the initial pressure was calibrated as 0.5 GPa. **b)** Stokes and anti-Stokes Raman spectra of 6L-MoS$_2$ under various pressures. The blue and red dashed lines indicated the vibrational frequencies of S$_{6,1}$, S$_{6,3}$, and LB$_{6,5}$ modes, respectively. **c)** Pressure-dependent Raman shifts (upper panel) and FWHM (lower panel) of S$_{6,1}$, S$_{6,3}$ and LB$_{6,5}$ modes of 6L MoS$_2$ nanoflake.

To investigate the *N*-dependence of low-frequency Raman shifts of MoS$_2$ under high pressure, we measured the Raman spectrum of MoS$_2$ nanoflakes with differnet *N* in DAC. The acquired Raman results of the lowest-frequency LB$_{N, N-1}$ and the highest-frequency S$_{N, 1}$ modes are displayed in Figure 3, as well as the DFT calculations. Apparently, for a MoS$_2$ nanoflake with the given *N*, both LB$_{N, N-1}$ and S$_{N, 1}$ modes monotonically display the evidently blueshifted vibrational energies (Figure 3a) and the boardened linewidths (Figures 3c) with pressure, similar with the results of 6L-MoS$_2$ in Figure 2. Interestingly, LB and S modes present the distinct Raman shifting rates with pressure, that is, d$\omega$(LB$_{N, N-1}$)/d$P$ (d$\omega$(S$_{N, 1}$)/d$P$) becomes smaller (larger) as the *N* increases. These exotic results was further confirmed by the theoretical calculations using both DFT (Figure 3b) and MD (Figure S6-S10) approaches. The quantitative anlysis will be discussed later in Figure 4. The Raman vibrations of bilayer MoS$_2$ still remained observable when the pressure exceeds 9.6 GPa, instead of the total vanishing in methanol-ethanol-water mixture, as shown in Figures S11 and S12. Futhermore, the FWHM of LB$_{N,N-1}$ mode was superior to that of S$_{N,1}$ mode, and both of them decreased with *N*, which is rather consistent with the lierature results.[32,35] As the applied pressure, for a given *N* of MoS$_2$, the FWHM of both LB$_{N,N-1}$ and S$_{N,1}$ modes became remarkably widened, implying the gradually distortted lattice or non-hydrostatic enviroment, as shown in Figure 3c.

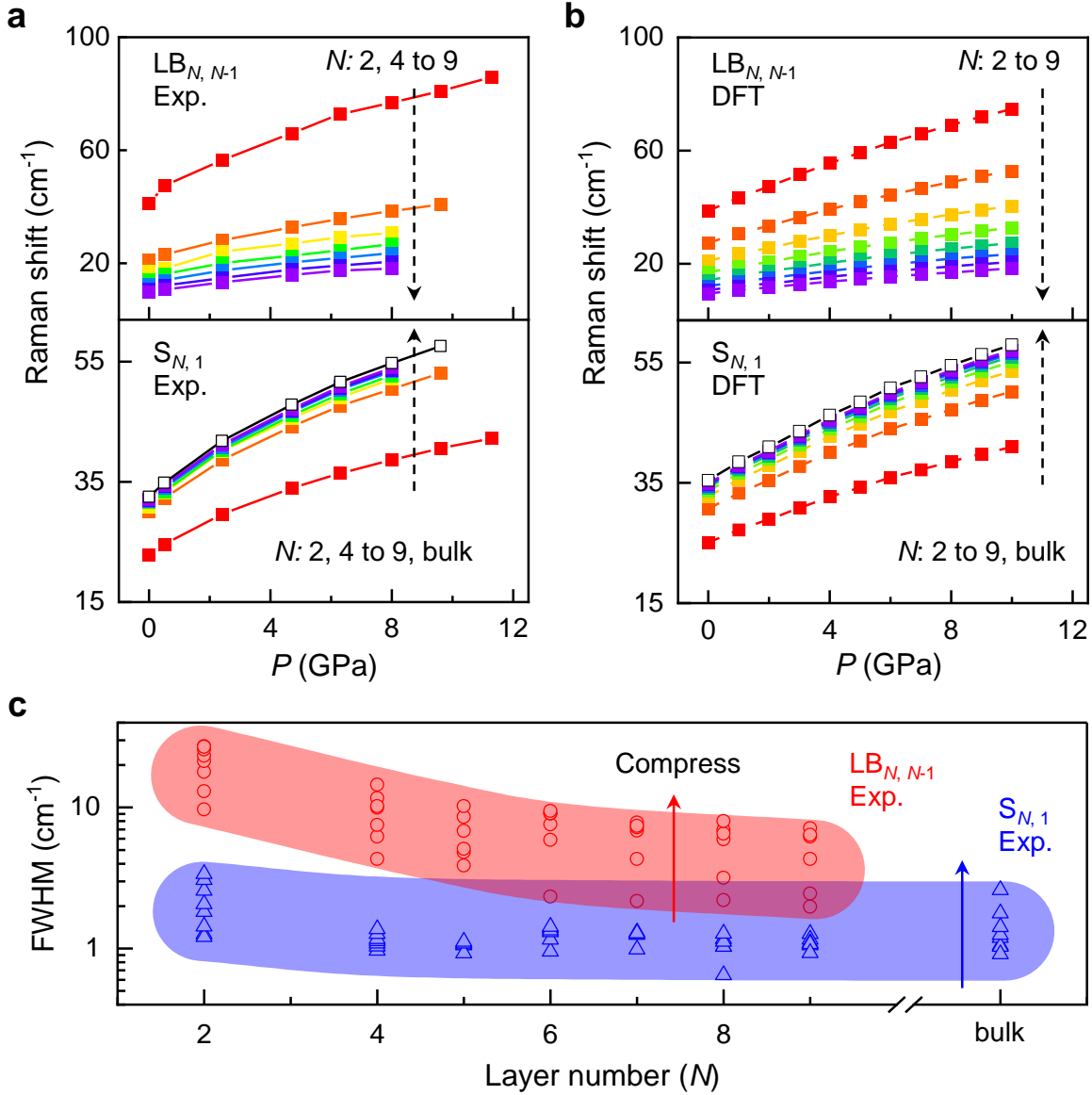

**Figure 3. Pressure-dependent Raman results of both LB and S modes of few-layer MoS$_2$ ($N$ up to 9) obtained by experiments (exp.) and theoretical calculations. a)** The measured Raman results of LB$_{N, N-1}$ (upper panel) and S$_{N,1}$ (lower panel) modes of MoS$_2$ with different $N$, which cover 2, 4 to 9, and bulk as indicated with dashed lines. **b)** The DFT calculated results of LB$_{N, N-1}$ (upper panel) and S$_{N,1}$ (lower panel) modes of few-layer MoS$_2$ with different $N$ ($N$ ranges from 2 to 9, and bulk). **c)** The experimental FWHM of LB$_{N, N-1}$ (red) and S$_{N,1}$ (blue) modes of MoS$_2$ with different $N$. The red and blue arrows indicate the increasing of pressure.

To better understand pressure effect on the interlayer interactions of $MoS_2$, we utilized MCM to thoroughly analyze the low-frequency Raman modes of $N$-layer $MoS_2$ under various pressures as shown in Figure 4a and S13, which allowed us to evaluate the S ($\alpha^{\parallel}$) and LB ($\alpha^{\perp}$) force constants, as well as the S modulus ($C_{44}$) and LB modulus ($C_{33}$). In terms of nanomechanics, we understand that $C_{33} = \alpha^{\perp} t$ and $C_{44} = \alpha^{\parallel} t$, where $t$ corresponds to the equilibrium distance between the neighboring layers of $MoS_2$ under the specific pressure condition.[14,20] As shown in Figure 4b, the elastic constant $\alpha^{\parallel}$ of S mode dramatically increased from $2.8 \times 10^{19}$ (0 GPa) to $8.4 \times 10^{19}$ (8 GPa) N m$^{-3}$. In contrast, the $\alpha^{\perp}$ of LB mode responds more quickly on pressure from $8.6 \times 10^{19}$ N m$^{-3}$ at 0 GPa to $31.4 \times 10^{19}$ N m$^{-3}$ at 8 GPa, owing to the more significant compression of interlayer distance under pressure. The derived $C_{33}$ and $C_{44}$ modulus presents the similar behaviors after calibrating the thickness of $MoS_2$ under pressure, consistent with our DFT calculations and literature results.[36] Notably, there exists a certain differences in force constant/modulus vs. pressure digram between experiments and simulations. These differences are reasonablly related to the different sample sizes and pressure conditions in experiments and simulations.

We further summarized all the d$\omega$/d$P$ data of LB and S modes for $MoS_2$ with different $N$, obtained by experiments and DFT calculations as shown in Figure 4c. It is clear that as $N$ increased, d$\omega$(S)/d$P$ gradually increased and then became saturated at about 6 layers, while d$\omega$(LB)/d$P$ conversely decreased and became saturated at ~9 layers. Moreover, when the $N$ was less than 4, the d$\omega$(LB)/d$P$ values were obviously larger than d$\omega$(S)/d$P$, however, the results are reversed when $N \geqslant 4$. This can be well explained by the MCM. For the $S_{N,1}$ mode, we know d$\omega(S_{N,1})$/d$P$ = [d$\omega$($S_{bulk}$)/d$P$]sin[$\pi(N-1)/2N$]. Since d$\omega$($S_{bulk}$)/d$P$ is kept as a constant, d$\omega$($S_{N,1}$)/d$P$ monotonically increases as the layer number $N$. Breathing mode $LB_{N, N-1}$ has the similar result, d$\omega(LB_{N, N-1})$/d$P$ = [d$\omega$($LB_{bulk}$)/d$P$]sin($\pi/2N$), and thus d$\omega(LB_{N, N-1})$/d$P$ decreases as $N$ increases. Morevoer, $C_{44}$ is more sensitive to pressure than $C_{33}$, leading to d$\omega$($LB_{bulk}$)/d$P$ > d$\omega$($S_{bulk}$)/d$P$. When $N = 2$, d$\omega(S_{2,1})$/d$P$ < d$\omega(LB_{2,1})$/d$P$ can be decuded. When $N$ approaches the infinity, d$\omega(LB_{N, N-1})$/d$P$ ~ 0 < d$\omega(S_{N,1})$/d$P$ ~ d$\omega$($S_{bulk}$)/d$P$. So there must be a crossover between d$\omega(S_{N,1})$/d$P$ and d$\omega(LB_{N, N-1})$/d$P$ at a finite $N$, well consistent with our experimental resutlts in Figure 4c. Alternatively, this phenomenon might be understood by different stress states of $MoS_2$ nanoflakes with different $N$ as well. [37] For the ultrathin $MoS_2$ ($N\sim2$), it mostly endures the uniaxial compression. For the thicker and bulk MoS$_2$, hydrostatic pressure dominates Raman responses of MoS$_2$ over stress conditons.

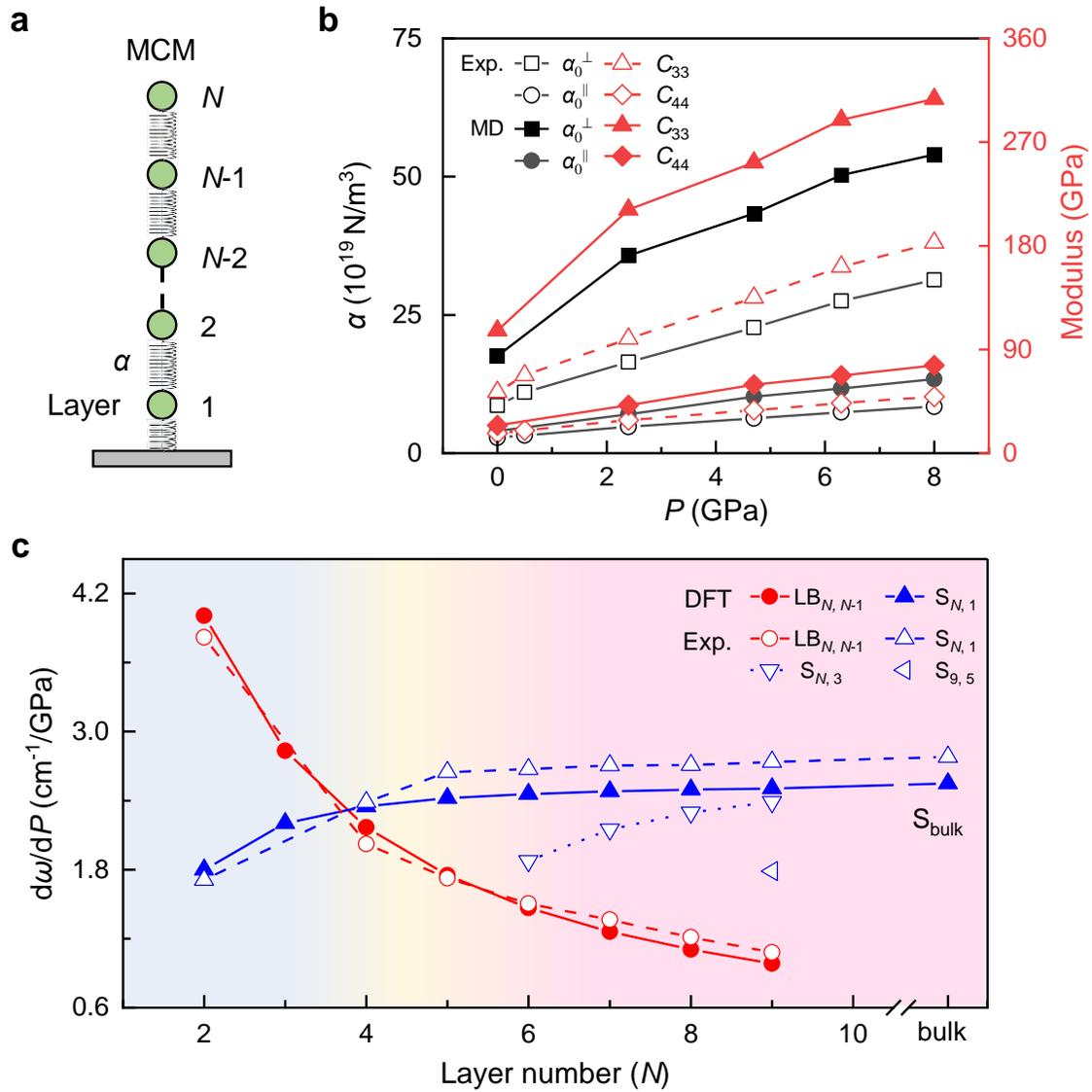

**Figure 4. Pressure-modulated elastic constants and lattice dynamics of few-layer MoS$_2$. a)** MCM of MoS$_2$ on a solid substrate. **b)** Evolutions of force constants ($\alpha^\perp$ and $\alpha^\parallel$) and modulus ($C_{33}$ and $C_{44}$) of LB and S modes of MoS$_2$ under high pressure. The solid and empty symbols denote the DFT and experimental results, respectively. The changes in the force constant and modulus under high pressure seem linear, leading to the nonlinear dependence of Raman scattering on pressure. **c)** $N$-dependence of Raman shifting rates of d$\omega$(LB)/d$P$ and d$\omega$(S)/d$P$ of MoS$_2$. Note that d$\omega$(LB)/d$P$ and d$\omega$(S)/d$P$ intersect when $N$ approaches to ~4. The empty and solid symbols denote the exp. and DFT results, respectively. The dahsed and solid lines are guides for the eyes.

In order to further elaborate the influence of external pressure on elastic constants of MoS$_2$, diatomic chain model (DCM) theory proposed by Wieting,[24,38] which considers both intralayer and interlayer interactions, was adopted to analyze our experimental data, as shown in Figure 5a. In this model (inset of Figure 5a), the shear and compressive force constants between the chalcogen planes of the neighboring layers are defined as $C_b^s$ and $C_b^c$, and between the molybdenum and chalcogen planes in a layer are $C_w^s$ and $C_w^c$, respectively. Based on Raman results of the in-plane $E_{2g}^1$ and $E_{2g}^2$ in Figure 5a, 3a and Note 2 in Supplementary Information, the values of $C_w^s$ and $C_b^s$ under each pressure can be deduced readily, as shown in Figure 5 and Figure S14. The values of $C_b^s$ rapidly increased with pressure, among which bilayer MoS$_2$ changes the slowest as shown in Figure 5b, and the change rate increases gradually with the increase of $N$ in Figure 5c. It is obvious that the slope values become saturated as the $N$ increases to 6, quantitatively consistent with the results of MCM mentioned above in Figure 4c. Similarly, the slope change of $C_w^s$ has the comparable trend with that of $C_b^s$. The linear slope d$C_b^s$/d$P$ of $N = 2$ and bulk MoS$_2$ were calculated as 0.27×10$^{-2}$ and 0.58×10$^{-2}$ cm$^{-1}$/GPa, respectively. In addition, the ratio of $C_b^s/C_w^s$ indicates the relative strength of interlayer and intralayer interactions, and thus the greater the ratio, the stronger the interlayer coupling. Figure 5d shows the calculated $C_b^s/C_w^s$ data of MoS$_2$ with different $N$. With the increase of the $N$, the $C_b^s/C_w^s$ ratio became larger, indicating that the interlayer interactions and the $N$ displayed the basically positive correlation. The values of $C_b^s/C_w^s$ gradually increases with pressure, but it remained less than 4.5% under ~8 GPa, indicating that the $N$L MoS$_2$ still retained the acceptable two-dimensional characteristic even under such high pressure.

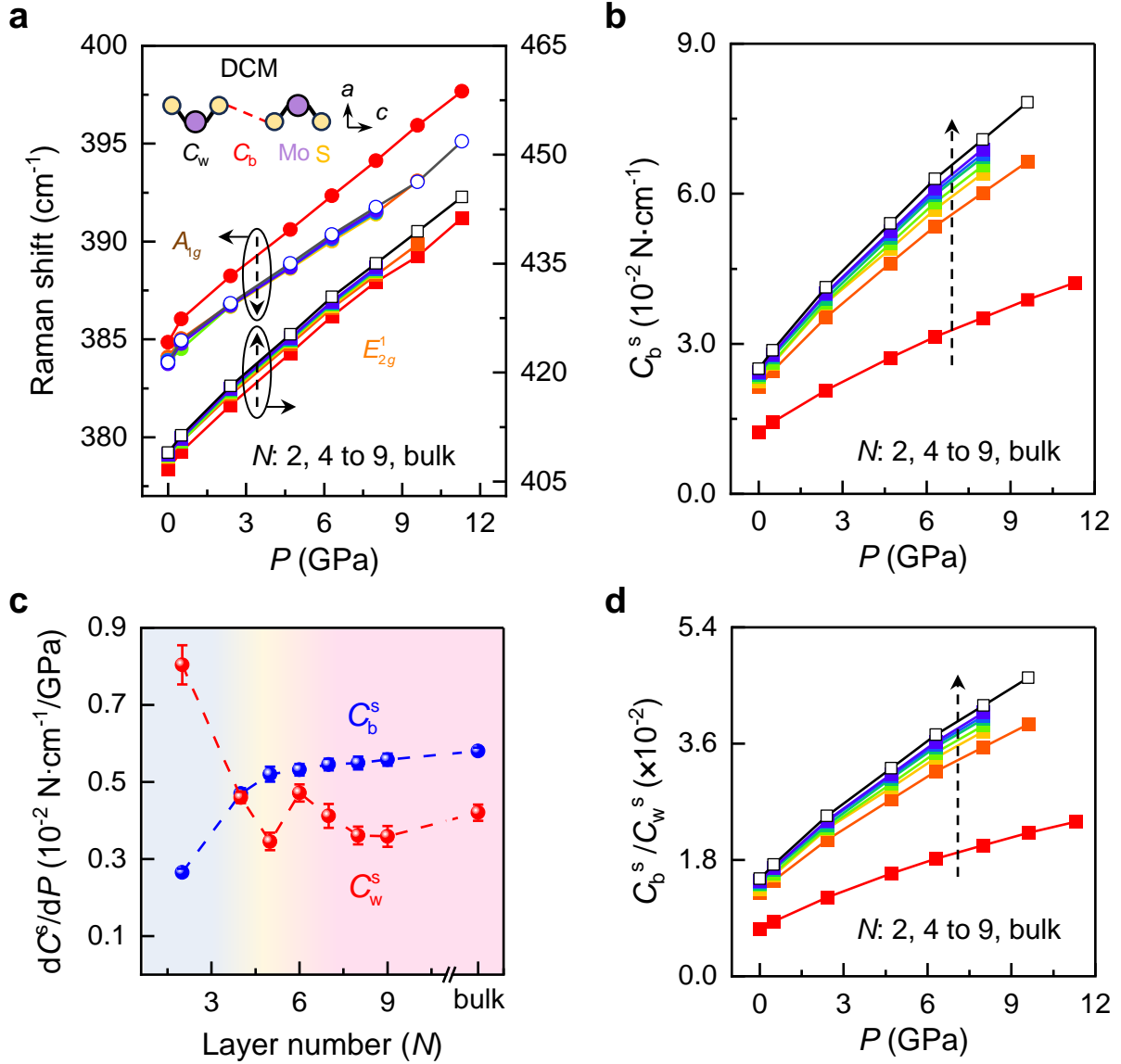

**Figure 5. DCM used to analyze the Raman vibrations and elastic constants of MoS$_2$ under high pressure. a)** High-frequency Raman modes $A_{1g}$ and $E^1_{2g}$ of MoS$_2$ with various $N$, and $N$ changed from 2, 4 to 9, and bulk. Insets shows the DCM of MoS$_2$. **b)** Pressure-dependence of shear force constant $C^s_b$ between the chalcogen planes of the neighboring layers with various $N$. **c)** The extracted d$C^s_w$/d$P$ and d$C^s_b$/d$P$ slops of MoS$_2$ as a function of $N$ under pressure. The $C^s_w$ is the shear force constant between Mo and S atomic planes in an individual layer. **d)** The obtained $C^s_b/C^s_w$ of MoS$_2$ with different $N$.

**CONCLUSION**

In conclusion, we systematically investigated the interlayer and intralayer interactions of 2D layered $MoS_2$ under extreme pressure by the ULF Raman spectroscopy, together with DFT calculations and MD simulations. The obtained $N$-dependent Raman results were analyzed through the MCM and DCM. Both interlayer and intralayer force constants gradually increased as pressure, with a more significant increment in the interlayer coupling owing to its rapid pressure response. Furthermore, it was found that the shifting rates of LB modes gradually decreased as $N$ became larger, dramatically different from the strengthened S modes. The current study is expected to offer novel insights into pressure-engineering of physical properties of 2D materials, and further promote their potential applications in nanomechanics and optoelectronics.

# EXPERIMENTAL METHODS

## Preparation of MoS$_2$ Nanoflakes and the Characterizations

Few-layer MoS$_2$ nanoflakes were prepared by using mechanically exfoliation approach from the commercial MoS$_2$ crystals with high quality (2D Semiconductor, USA). Polydimethylsiloxane was utilized as the template, instead of the conventional Scotch tape to exclusive any tape residual, and then MoS$_2$ nanoflakes were directly transferred onto the culet of diamond. The entire dry-transfer process can guarantee the clean MoS$_2$ surface without any contamination or dirt. Layer number of few-layer MoS$_2$ nanoflakes was exactly determined by combining optical contrast and low-frequency Raman spectrum.

## High-Pressure Raman Spectroscopy

We used a symmetric DAC with a ~400 μm culet to generate high pressure on MoS$_2$ nanoflakes. A T301 gasket with the central ~160 μm hole drilled by laser milling was used as the sample chamber. The pressure was calibrated by the shifted fluorescence peak of the tiny ruby ball, placed near to the MoS$_2$ nanoflakes in DAC. The methanol-ethanol-water (16:3:1) mixture was used as pressure transmission medium, in order to compare the hydrostatic pressure effect on Raman spectrum of MoS$_2$.

All Raman spectrum were acquired through our homemade Raman system, which is based on the SmartRaman confocal-micro-Raman module (Institute of Semiconductors, Chinese Academy of Sciences) and equipped with the Horiba iHR 550 spectrometer and 532 nm laser as excitation source.Importantly, the laser power was optimized to be low enough, in order to avoid the possible over-heating effect. A single-longitudinal-mode laser with narrow linewidth and high stability (MSL-III-532), and a narrow linewidth (better than 10 cm$^{-1}$) bandpass filter (BPF) based on the volume Bragg grating (VBG) technique was used to detect the low-frequency Raman modes.[33] Three VBG-based notch filters (BNF) (bandwidth: ~8-10 cm$^{-1}$) with high transmittance (up to 80%-90% dependent on the laser wavelength) and an optical density (OD) 3-4 were strategically used to suppress the strong Rayleigh background, and the BPF was utilized to remove any plasma line of laser source. Moreover, two mirrors behind the reflecting filter (BPF) were used to align the laser beam to the center of the first BNF, as shown in Figure S3. The angles of three BNFs were precisely tuned to block the Rayleigh line up to 10$^{-9}$-10$^{-12}$. An iHR 550 spectrometer was equipped to detect the Raman scattering signals.

**DFT Calculations**

DFT calculations were performed by the Vienna Ab initio Simulation Package.[39,40] The projector-augmented wave method[41] was adopted to describe the core-valence interaction. The exchange-functional was treated by the generalized gradient approximation of the Perdew-Burke-Ernzerhof functional.[42] The energy cut-off for the plane wave basis expansion was set to 500 eV and the convergence criterion of geometry relaxation was set to 5 meV/Å. The self-consistent calculations applied a convergence energy threshold of $10^{-5}$ eV. A $12 \times 12 \times 1$ Monkhorst-Pack $k$-point grid[43] was used to sample the Brillouin zone of the unit cell of bulk $MoS_2$. The van der Waals interaction was included through Grimmes's DFT-D3 method.[44] The phonon modes were calculated using the finite displacement method.

**MD simulations**

To evaluate the dynamic characteristics of multi-layer $MoS_2$ under pressure, we performed a series of large-scale MD simulations for the vibration of multilayer $MoS_2$ under different pressures. All MD simulations were performed via the large-scale atomic/molecular massively parallel simulator (LAMMPS)[45]. We constructed 2- to 6-layer 2H-$MoS_2$ samples with the same in-plane dimensions of $8.19 \times 8.18$ nm$^2$ and an interlayer spacing of 0.62 nm. Throughout simulations, periodic boundary conditions were imposed along in-plane directions. The reactive force field (ReaxFF) potential[46] was adopted to describe the interatomic interactions of Mo and S atoms. Such interatomic potential includes the covalent bonding interaction, and non-bonded van der Waals and Coulomb interactions[46]. This potential can accurately describe the thermodynamic, structural and mechanical properties and formation energies of various defects of single- and multiple-layer $MoS_2$ systems[46]. All simulated samples were equilibrated by initial energy minimization and subsequent free relaxation at 300 K for 100 picoseconds. During free relaxation, the temperature kept at 300 K and the pressures along in-plane directions were controlled as zero by using an isothermal-isobaric ensemble[47]. After equilibration, the simulated samples were imposed by the hydrostatic pressure. The pressures along in-plane directions were controlled to reach the target pressure (0, 2.4, 4.7, 6.3, and 8.0 GPa) via an isothermal-isobaric ensemble[47], while the pressure along out-of-plane direction was controlled to achieve the desired pressure value by adjusting the positions of a pair of virtual walls via a proportional-integral-derivative controller. This pair of virtual walls are parallel to the $MoS_2$ layers, and located on the top and bottom of simulated samples, respectively. The interaction energy between walls and atoms in the simulated samples is described by a repulsive-only harmonic spring potential,

$$E = \varepsilon(r - r_c)^2, \qquad r < r_c \tag{1}$$

where $\varepsilon$ is the spring constant and $r_c$ is the cutoff distance for the given wall. The values of $\varepsilon$ and $r_c$ were set as 0.13 eV/Å$^2$ and 8.0 Å, respectively. After the hydrostatic pressure reached the target value, the simulated samples were relaxed for 250 ps under the target pressure. The temperature kept at 300 K via an isothermal-isobaric ensemble[47]. Subsequently, we monitored the centroid coordinates of each MoS$_2$ layer at an interval of 1 fs for 50 ps and obtained the evolution curve of vibration displacement with time of each layer. Then we performed fast Fourier transform on the vibration displacement-time curves to make the spectrum and modal analyzes on the variation of simulated samples under hydrostatic pressure.

## ASSOCIATED CONTENT

**Supporting Information**

The Supporting Information is available free of charge online.

Normal mode displacements for the interlayer LB and S modes in few-layer $MoS_2$; Schematic diagram of our homemade Raman system for ultralow frequency measurements; MD and DFT calculated Raman results of 2D $MoS_2$ under pressure; The experimental Raman spectrum of $MoS_2$ with various thicknesses under pressure; Notes on the MCM and DCM to analyze the lattice dynamics of $MoS_2$ under pressure.

**Conflict of Interests**

The authors declare no competing financial interests.

**Data Availability**

The data that support the findings of this study are available from the corresponding author upon reasonable request.

**Author Contributions**

Y.C. and W.X. conceived this research project and designed the experiments. G.D. designed and established the ultra-low frequency Raman spectroscopy with the help of L.L.Z. and S.F. G.D., L.L.Z., W.H., J.L., Y.D., J.M. and T.Z. prepared the nanoflake samples and further performed the *in situ* Raman characterizations and high-pressure experiments. S.L. and X.L. performed the atomistic simulations. J.H. and J.K. performed the DFT calculations and theoretical analyses with J.Z. and G.D. Y.C. and G.D. wrote the manuscript with the essential inputs of all co-authors. All authors have given approval of the final manuscript.

**Acknowledgements**

This work was financially supported by the National Natural Science Foundation of China (grant numbers 52072032, 12090031, 12074029, 11991060, and 12325203), and the 173-JCJQ program (grant No. 2021-JCJQ-JJ-0159).

**For Table of Contents Only**

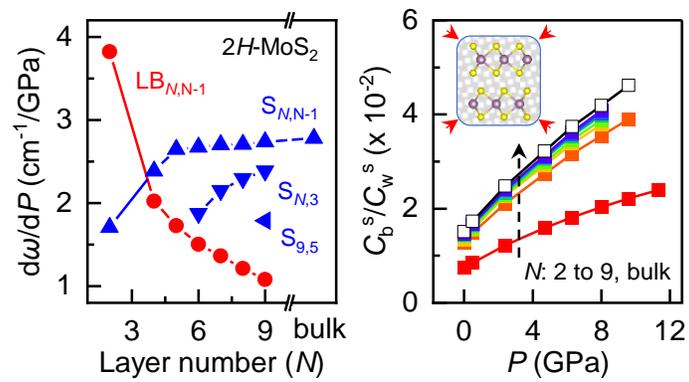